# Localised vibrations in superconducting YBCO revealed by ultra-fast optical coherent spectroscopy

Fabio Novelli,[1,2] Gianluca Giovannetti,[3] Adolfo Avella,[4,5] Federico Cilento,[2] Luc Patthey,[6] Milan Radovic,[6,7] Massimo Capone,[3] Fulvio Parmigiani,[1,2,8] and Daniele Fausti[1,2,*]

[1] Department of Physics, Università degli Studi di Trieste, 34127 Trieste, Italy

[2] Sincrotrone Trieste SCpA, 34127 Basovizza, Italy

[3] CNR-IOM Democritos National Simulation Center and Scuola Internazionale Superiore di Studi Avanzati (SISSA), Via Bonomea 265, 34136 Trieste, Italy

[4] Dipartimento di Fisica, E.R. Caianiello - Unità CNISM di Salerno, Università degli Studi di Salerno, I-84084 Fisciano (SA), Italy

[5] CNR-SPIN, UoS di Salerno, I-84084 Fisciano (SA), Italy

[6] SwissFEL, Paul Scherrer Institute, CH-5232 Villigen PSI, Switzerland

[7] Institut de la Matiere Complexe, EPF Lausanne, CH-1015 Lausanne, Switzerland

[8] International Faculty, University of Cologne, Germany

∗daniele.fausti@elettra.eu

The interaction between phonons and high-energy excitations of electronic origin in cuprates and their role in the superconducting mechanisms is still controversial. Here we use coherent vibrational time-domain spectroscopy together with density functional and dynamical mean-field theory calculations to establish a direct link between the c-axis phonon modes and the in-plane electronic charge excitations in optimally doped YBCO. The non-equilibrium Raman tensor is measured by means of the broadband "coherent-phonon" response in pump-probe experiments and is qualitatively described by our model using DFT in frozen phonon approximation plus single band DMFT to account for the electronic correlations. The major outcome of our experimental and theoretical study is to establish the link between out-of-plane copper ions displacements and the in-plane electronic correlations, and to estimate at few unit cells the correlation length of the associated phonon mode. The approach introduced here could help revealing the complex interplay between fluctuations of different nature and spatial correlation in several strongly-correlated materials.

## I. INTRODUCTION

The role played by phonons in cuprates superconductivity is still controversial. While a purely phononic mechanism can hardly account for the superconducting state [1,2], strong anomalies are found in the phonon sub-system upon entering the superconducting phase [3,4]. Furthermore, non-standard fingerprints of electron-phonon interaction have been observed as a result of the interplay with strong electron-electron correlations [5]. A common feature of the cuprates is that the onset of the superconducting phases is accompanied by large changes in the optical properties up to energies of order of few eV [6–9], i.e. about one order of magnitude larger than typical superconducting gaps (10-100 meV). This is in striking contrast with a BCS scenario in which the variations of the optical properties at the superconducting phase transition are limited to energies of the order of the superconducting gap. Such an unusual behaviour reveals the strong interconnection between high-energy processes, mainly controlled by electron-electron correlations, and the low energy excitations relevant in the pairing mechanism of the cuprates [10].

The nature of the electronic excitations in doped cuprates and other strongly correlated materials is highly debated. Qualitatively, and neglecting for simplicity important dressing phenomena [11], the bare carriers properties are best understood at doping extremes. For undoped compounds the charge is localised at the copper sites by carrier-carrier correlations, while for strongly overdoped materials a quasi-Fermi-liquid behaviour emerges and the excitations are delocalised [12]. At intermediate doping one could think that the low-energy fluctuations have a mixed character or, equivalently, that both localised and delocalised excitations are present. Time domain "coherent-phonon" experiments may offer a new ground to access the spatial and temporal correlation of phonon modes in complex materials. For the sake of clarity let us consider a cartoon of two complementary coherent phonon modes, here identified as 'delocalised' and 'localised', and shown in Fig.1a and Fig.1c, respectively. Broadly speaking, one expects that for a 'delocalised' mode a larger excitation density results in a larger displacement of the ions from their equilibrium positions (Fig.1b) and eventually leading to anharmonic effects characterised by shifted spectral responses. Conversely, if the atomic displacement is 'localised' on a few lattice sites, the vibrational response ought to be linear until all sites are excited (Fig.1d) and, after that, it should saturate without frequency shifts.

Here we report on a novel approach, based on time-domain broadband spectroscopy, density functional theory (DFT), and dynamical mean field theory (DMFT)[13], to address the interplay between vibrational modes and the high-energy electronic response in the cuprates. By performing pump-probe measurements as a function of temperature and excitation density, we are able to map the characteristic responses of specific low-energy vibrations involving almost-pure barium (Ba) and copper (Cu) c-axis displacements [14]. In agreement with previous results [14-21], the photo-excited amplitude associated to the Ba mode maps the onset of the superconductivity while the amplitude of the oscillation associated to the Cu displacement seems independent from the superconducting phase. Here we perform an intensity-dependent study and, by comparing experimental and theoretical results, we are able to measure the correlation length of the atomic displacements triggered by visible photon absorption.

**II. METHODS**

Extensive pump-probe measurements have been performed on 80 nm thick optimally doped YBCO films grown by pulsed laser deposition on STO(001)[22] ($T_C$ = 88 K) as a function of temperature, T, and pump fluence, $\Phi$. The temperature was set in a liquid He cryostat between T = 5 K and T = 120 K, while the excitation pump density was tuned within $\Phi$ = 20 $\mu J/cm^2$ and $\Phi$ = 180 $\mu J/cm^2$. The pump pulses are ~80 fs long with a central energy $E_{pump}$ = 0.95 eV. The time and temperature-dependent reflectivity $\Delta R/R$ of optimally doped YBCO was measured on a large spectral region with broadband white-light probes generated in a sapphire crystal (1.2 eV < $E_{probe}$ < 3.1 eV with a fluence lower than 5 $\mu J/cm^2$, see Supplementary Figure 1).

The measurements at a pump fluence $\Phi$ = 20 µJ/cm$^2$ and at probe energy $E_{probe}$ = 2.2 eV show oscillating components which we associate to "coherent" vibrational excitations (Fig. 2a). These oscillations, which are the main target of this work, are superimposed to a decaying function of time, that is labeled as "incoherent" in the following. In Fig. 2b we show the incoherent part at the fixed pump-probe delay of 0.3 ps versus temperature. In Fig.2c we show the effect of the increased fluence on the relative variation of the reflectivity $\Delta R/R$ at 2.2 eV probe energy and at 0.3 ps pump-probe delay (T = 5 K).

The coherent contribution is obtained from the measurements, as outlined in Fig. 3. First we subtract the incoherent part by fitting $\Delta R(E_{probe},t)/R$ with a multi-exponential decay convoluted with a step function at each $E_{probe}$. Then we obtain the difference between the data and the incoherent part, as shown in Fig. 3a for T = 5 K and pump fluence of 20 µJ/cm$^2$. This residual difference contains only the oscillating part and is subsequently Fourier transformed (Fig. 3b-d) to give a Fourier amplitude depending on frequency and probe photon energy. Two oscillations can be clearly identified with central frequencies at about 3.5 THz and 4.5 THz, respectively.

## III. ANALYSIS

The changes of the optical properties of the cuprates at the onset of the superconducting phase [6–9] are particularly evident in pump-probe experiments [23–28]. Actually, when pump-driven charge excitations modify the electron distribution the condensate is suddenly de-stabilised [29–31]. In such conditions, the anomalous connection between the formation of the condensate and the high-energy optical transitions leads to large variations of the time-resolved reflectivity $\Delta R(E,t)/R$ in the visible range[32–38]. This is striking in e.g. Fig.2b, where the formation of the superconducting phase as a function of temperature can be clearly mapped by the high-energy reflectivity response.

Also the fluence-dependent response is linked to the superconducting phase. In fact at low fluence $\Delta R/R$ is negative and increases in absolute value as the fluence grows (Fig.2c). At the critical value $\Phi_{SC} \sim$ 50 µJ/cm$^2$ a sharp kink is observed, followed by an increase of $\Delta R/R$ which eventually turns positive for larger fluences. This trend is a well-known signature of the photo-induced collapse of the superconducting gap observed in various compounds of the cuprate family[39–41].

Also the coherent part of the time domain response is sensitive to the onset of superconductivity, because it consists of one single mode (4.5 THz) above the critical temperature, while a second oscillation frequency (3.5 THz) appears below $T_C$ (Fig.2d). These frequencies, already observed in other Raman[15,16] and pump-probe experiments[17–20], correspond to two vibrational modes of $A_{1g}$ symmetry involving almost pure barium (Ba) and copper (Cu) out-of-plane vibrations, respectively[14,16,21] (sketched in Fig. 2e). In good agreement with previous Raman studies at equilibrium[15,16], the spectral response associated with these oscillations is reported in Fig. 3c and Fig. 3d. Further details on the fitting procedure employed to obtain the coherent contributions can be found in Supplementary Figure 2.

## IV. COHERENT RESPONSE AT 4.5 THz

### IV.A. Experimental results

The coherent part at 4.5 THz, associated with almost pure Cu oscillations along the c-axis, has been extracted from all the T- and $\Phi$-dependent pump-probe measurements following the procedure outlined in Fig. 3 and described in the previous section. The results for this Cu mode are shown in Fig. 4. As it can be seen by inspecting Fig. 4a, where the average spectral response between 2.1 eV and 2.3 eV is reported, this vibration seems independent on temperature. The spectral amplitude of the Cu mode is shown in Fig. 4b as a function of pump fluence. In this case we found an initial roughly-linear increase followed by saturation above about $\Phi_{crit}$ = 120 μJ/cm$^2$. The normalised spectral responses versus temperature (Fig. 4c) and fluence (Fig. 4d) are scarcely affected.

Two important conclusions can be drawn for the coherent response at 4.5 THz. Firstly, the Cu mode seems not coupled to the superconducting phase. Actually, the temperature behaviour (Fig. 4a) is smooth through all the superconducting phase transition at $T_C$ ~ 88 K. It is also known that for a critical fluence of about $\Phi_{SC}$ ~ 50 μJ/cm$^2$ (Fig. 2c) the superconducting phase is photo-melted and the corresponding transient reflectivity slope changes. However, the fluence dependence of the 4.5 THz vibration displays none of these features (Fig. 4b). Furthermore, the fluence response of the Cu mode is simply consistent with the response expected for a localised displacement of the ions[19] (see Fig. 1c and Fig. 1d).

**IV.B. Model results**

In order account for the energy-dependent coherent response of the Cu mode outlined in Fig. 4, it is necessary to consider the strong electron-electron correlations which are responsible for the insulating state of the parent compounds and for the distribution of spectral weight in the visible range[42]. This is implemented combining the DFT band structure (dashed line in Fig.5a), calculated with the generalized gradient approximation (PBE) using Quantum Espresso[43] and the experimental lattice parameters of Ref.[44], with a DMFT treatment of the Coulomb interaction. We employed a Wannier projection[45] including the $d_{x^2-y^2}$ orbital and an on-site Coulomb repulsion on the copper sites of U = 4.8 eV, and proceeded with the exact diagonalisation as the impurity solver with 8 sites in the bath.

In Fig. 5a we compare the DMFT density of states with the uncorrelated results from DFT. The prime effect of the electronic correlation is to shift spectral weight to the lower (LHB) and to the upper Hubbard band (UHB), which coexists with a low-energy structure corresponding to itinerant carriers with a renormalized bandwidth (FL). In Fig. 5b we show the real part of the optical conductivity $\sigma_1(\omega)$ calculated from the band-structure of the DMFT. As underlined by the grey shaded areas in Fig. 5a and Fig. 5b, $\sigma_1(\omega)$ in the visible region is dominated by transitions between the LHB and the electronic states close to the Fermi level. This observation enucleates the origin of the conductivity variations up to energies as high as few eV following the opening of a superconducting gap of the order of few tens of meV.

Now we rationalize the amplitude of the coherent response involving Cu displacements by means of a differential approach. Initially we argue that the coherent amplitude can be qualitatively described by the difference between $\sigma_1(\omega)$ calculated with the ions at the equilibrium positions, and the one obtained after displacing the ions along the phonon eigenvector. Several deformations have been used and the calculation results for two significant ones, **d$_1$** and **d$_2$**, are shown in Fig. 5c. These deformations are obtained by artificially moving the ions along the eigenvectors of the 4.5 THz $A_{1g}$ phonon by 1% and 2%, respectively[16]. These displacements have been chosen because of the qualitative agreement between the spectral responses as a function of fluence: when the deformation increases from 1% to 2%, the spectral response associated with the Cu mode gets slightly broader as for larger-fluence experimental observations.

Fig.5c displays the relative variations of the optical conductivity driven by the out-of-plane displacement of Cu ions. These differences are compared with the coherent Cu amplitude as a function of the probe energy (Fig.5d is reproduced from Fig.4d for display purposes). We argue that this comparison is reasonable also below $T_C$ because the model does not account for the superconducting phase and the experiments reveal that the Cu mode is not coupled to it (see Fig. 4). Such a comparison shows that a proper treatment of the on-site Coulomb interaction on copper allows to account, on a qualitative level, for the energy dependence of the coherent-phonon response even in the single-site DMFT approximation, without including other bands, longer-range interactions, and explicit phonons couplings.

### IV.C. Correlation length

The correlation length of the localised lattice deformation associated with Cu c-axis displacements can be estimated both from the experimental results and by comparing measurements with the DMFT calculations. Both approaches give consistent results, as explained in the following.

*IV.C.1. Correlation length from the experiments*

In order to estimate the correlation length of the Cu mode we start with the number of photons per unit cell at the critical fluence, $\Phi_{crit}$ = 120 µJ/cm$^2$, marking the appearance of saturation effects (Fig. 4b). Considering that the penetration depth of YBCO at $E_{pump}$=0.95 eV/ph (energy per photon) is about 320 nm[46] and that the film is 80 nm thick, we estimate a critical pump energy density δ ~ 2.22·10$^{-5}$ ph/Å$^3$ (photon per cube Ångström). Given the unit cell (u) volume of 174 Å$^3$/u[47], the critical fluence corresponds to 3.86·10$^{-3}$ ph/u or, equivalently, that there is one excited site in about 250 unit cells. Assuming for simplicity a spherical volume distribution, we can simply take its diameter as the correlation length. This diameter is about 8 u and hence the Cu mode should be considered as a 'localised phonon' according to the cartoon introduced in Fig.1.

*IV.C.2. Correlation length from model calculations*

We can estimate the correlation length of the 4.5 THz coherence by comparing the results of pump-probe experiments and model calculations. We note that the amplitude of the spectral response is strongly overestimated by the DMFT calculations. In fact, while the model predicts $\Delta\sigma_1/\sigma_1$ ~ 0.1 at the probe energy of 2.2 eV, the coherent part of $|\Delta R(E=2.2\ eV,t)/R|$ is roughly 3·10$^{-4}$ for $\Phi_{crit}$ = 120 µJ/cm$^2$. The difference between model and experiment comes from the fact that the calculations are performed by imposing self-consistent conditions to the phonons. As in the calculations all sites are excited by the laser pump pulses, we obtain $\Delta\sigma_1/\sigma_1$ ~ 0.1 after considering ~1 ph/u (Fig. 1a,b) and averaging the response between 2.1 eV and 2.3 eV of probe energy. This is clearly not the case for the experiment, and we can estimate the number of photo-excited unit cells by taking the ratio $(\Delta R/R)_{experiment}/(\Delta\sigma_1/\sigma_1)_{theory}$ ~ 3·10$^{-4}$/0.1 ~ 3·10$^{-3}$ ph/u. This is very close to what has been obtained in the previous paragraph by analysing the experimental results alone, and allows to obtain a correlation length of ~ 9 u for the Cu mode at 4.5 THz. Even though this rather simplified calculation cannot be expected to reproduce a quantitative correspondence to experimental data, this result additionally supports the 'localised-phonon' origin of this low-energy coherent response[19].

### V. COHERENT RESPONSE AT 3.5 THz

As we discussed previously, the Cu mode (4.5 THz) is weakly unaffected if we increase the base temperature of the sample (Fig. 4a). On the contrary the Ba mode, associated with the coherent response at 3.5 THz, is linked to the superconducting phase[14-21]. The Ba coherent-phonon response as a function of probe energy, temperature, and fluence, is shown in Fig. 6. In Fig. 6a (Fig. 6b) the temperature (fluence) dependence clearly indicates that the coherent response at 3.5 THz is strongly coupled with the superconducting phase: it disappears at T > $T_C$ (Fig. 6a) and is strongly perturbed approaching the critical fluence (Fig. 6b) identified by the transient reflectivity measurements ($\Phi_{SC}$ ~ 50 μJ/cm$^2$, Fig. 2c). Accordingly, the normalised temperature-dependent spectral response in Fig. 6c is below the noise level for T > $T_C$ while the spectral response of the coherent fluctuations at 3.5 THz are strongly shifted at $\Phi > \Phi_{SC}$.

The effect of an increased pump fluence on the Ba mode does not mirror the effect of temperature. While the temperature essentially washes out the 3.5 THz coherent response without significantly affecting its spectral response, an increased pump intensity leads to a marked shift of the coherent vibrational response (Fig.6d). This is suggestive of a non-thermal character of the pump-driven excitations. The thermally-driven melting of the superconducting phase arises from a large increase of excitations around the nodal points of the superconducting gap, where low-energy states are available. On the other hand, the light-driven closure of the superconducting gap is mainly due to the transient population of excitations around the anti-nodal regions, in agreement with recent time-domain photoemission experiments[48,49]. Taking into account that the apical oxygen in YBCO is excited in a pump-probe experiment within 150 fs, and that such timescale is faster than the quasiparticles thermalization time[50], we can interpret our findings in terms of an effective momentum-dependent light-driven dynamics of the superconducting gap. Further theoretical studies in the superconducting state including out-of-plane degrees of freedom are needed to substantiate this intriguing scenario.

## VI. CONCLUDING REMARKS

In summary, we addressed the interaction mechanism between the low-energy off-plane vibrations and the high-energy in-plane electronic excitations in superconducting cuprates by means of time-resolved ultra-fast reflectivity measurements, density functional theory, and dynamical mean field theory. The broadband optical probe allowed for the full dynamical characterization of the Raman tensor of c-axis Barium and Copper vibrations. We revealed a coherent response at 4.5 THz, associated with a Cu off-plane vibration, which seems independent of the superconducting phase. A minimal model including realistic electronic structures and correlations could both reproduce qualitatively the spectral response of the 4.5 THz coherence and allow us to estimate its correlation length in optimally doped YBCO. On the contrary, the amplitude of the vibrational mode associated to Ba c-axis vibration is strongly sensitive to the onset of superconductivity. Our findings reveal that, upon a light-induced non-adiabatic manipulation of the superconducting phase [51], the amplitude of the Ba vibrational mode follows closely the response of the condensate. The anomalous shift of its Raman tensor provides a direct experimental link between the photo-dynamics of high-energy excitations and the onset of thermodynamic phases in the cuprates.

Finally, the approach introduced here shows that the hierarchy between the onset of anharmonicity and saturation could be used to access the spatial correlations of vibrational modes and might pave the way to novel, all-optical studies of the spatiotemporal fluctuations of the correlated excitations in the cuprates and other exotic materials.

**VI. ACKNOWLEDGEMENTS**

D.F., M.C., F.N., F.C. and F.P. acknowledge support by the European Union under FP7 GO FAST, grant agreement no. 280555. D.F acknowledges the support of the Italian Ministry of education (MIUR), through the grant# RBSI14ZIY2, financed under the Scientific Independence of young Researchers (SIR2014) program. M.C. and G.G. are financed by the European Research Council through the ERC Starting Grant n.240524 SUPERBAD.

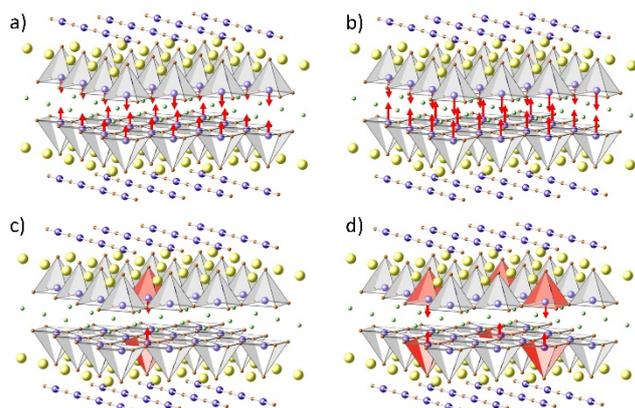

FIG. 1. Cartoon of 'delocalised' (a) vs 'localised' (c) c-axis phonon modes. Blue, red, yellow, and green spheres represent Copper, Oxygen, Barium, and Yttrium ions, respectively. Increasing the excitation density in a pump and probe experiment is expected to drive a larger atomic displacement for a 'delocalised' phonon mode (b), while an increased density of similarly displaced sites for a 'localised' phonon (d).

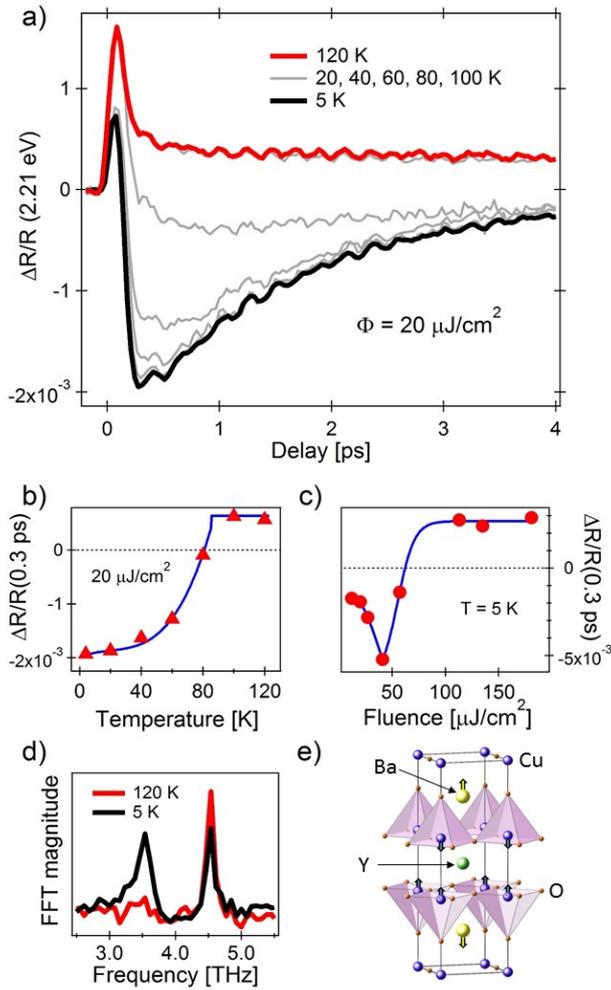

FIG. 2. (a) Transient reflectivity representative of the normal (red) and superconducting phase (black) at $E_{probe}$=2.2 eV for low pump fluence. (b) Relative variation of the 'incoherent' part of the reflectivity for a pump-probe delay of 0.3 ps as a function of temperature, for $\Phi$ = 20 μJ/cm². (c) Same quantity as (b) but versus the pump fluence at T = 5 K. In both (b) and (c) the dotted line marks zero, which is the result one would expect for no pump-induced changes to the reflectivity. (d) Amplitude of the 'coherent' contribution to the transient reflectivity as a function of frequency for T=5 K and T=120 K at low fluence. (e) Sketch of the displacements associated with the Ba (3.5 THz) and Cu (4.5 THz) ions.

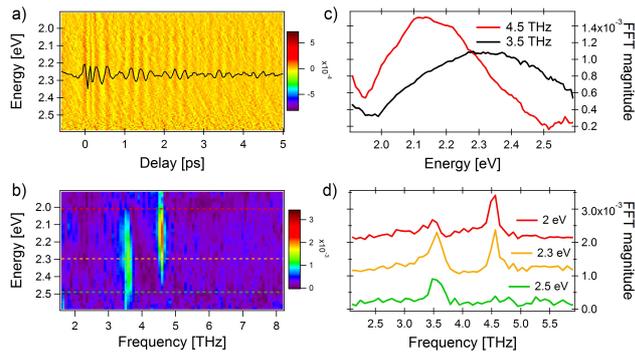

FIG. 3. (a) Amplitude of the coherent transient reflectivity (colours) at different probe energies, at T = 5 K, and for Φ = 20 μJ/cm² pump fluence. The black line is a representative cut at $E_{probe}$ = 2.2 eV. (b) Amplitude (colours) in the frequency-domain obtained by Fourier transforming the time traces in (a). (c) Amplitude of the two relevant vibrational modes versus probe energy. (d) Representative traces (dashed lines in (b)) with an offset added for clarity.

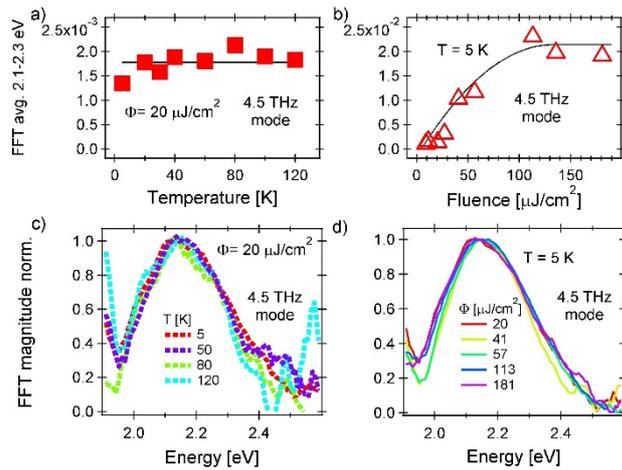

FIG. 4. a) Temperature-dependent coherent response at 4.5 THz averaged at probe energies within 2.1 eV and 2.3 eV. b) Fluence-dependent coherent response at 4.5 THz. c) Spectrum of the 4.5 THz oscillation as a function of temperature, normalised for the sake of comparison. d) Normalised fluence-dependent coherent response at 4.5 THz and T = 5 K.

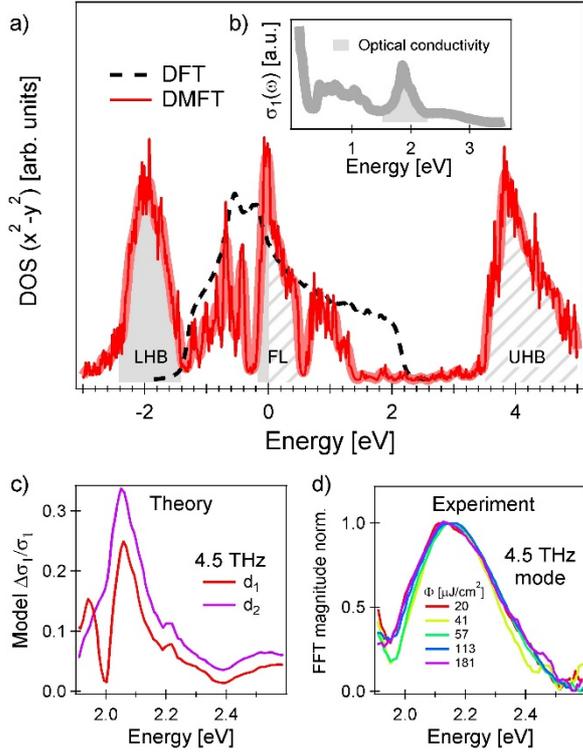

FIG. 5. (a) Density of states obtained in DMFT (red), compared with the DFT result (dashed black). (b) DMFT optical conductivity $\sigma_1(\omega)$. (c) Calculated $\sigma_1(\omega)$ changes in the visible range, driven by the vibrational mode at 4.5 THz for two exemplary displacements amplitudes ($d_1$ and $d_2$). (d) Energy-dependent amplitude of the Cu coherent response measured in pump-probe experiments.

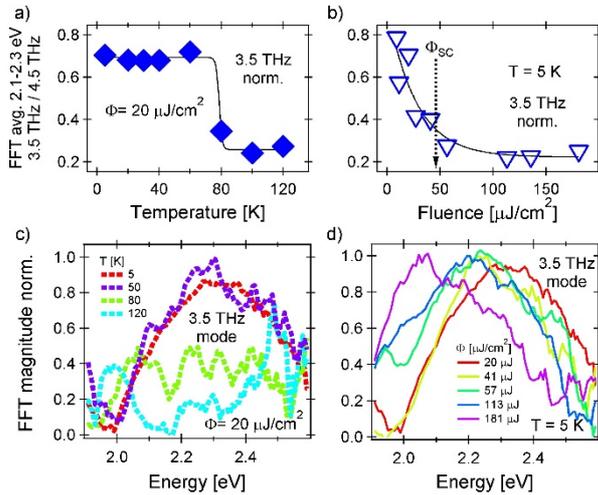

FIG. 6. (a) Amplitude of the coherent response at 3.5 THz, associated with the Ba mode, averaged between 2.1 eV and 2.3 eV and normalised on the Cu response (Fig. 4a). (b) Similarly normalised Ba coherence versus fluence. In (c) we show the spectral response as a function of temperature, and in (d) the one versus pump fluence. In panels c) and d) most spectra maxima are normalised to one to allow for a visual comparison of the changing energy dispersions. This is done properly for all curves except the ones at high temperature in panel c) (green and cyan) where the 3.5 THz mode is absent and the corresponding spectral functions are below the noise level.